\documentclass[twocolumn,aps,floatfix,superscriptaddress]{revtex4-2}
\usepackage[T1]{fontenc}
\usepackage[utf8]{inputenc}
\usepackage{units}
\usepackage{mathrsfs}
\usepackage{mathtools}
\usepackage{amsmath}
\usepackage{multirow,enumitem}
\usepackage{amssymb}
\usepackage{graphicx}
\usepackage{esint}
\PassOptionsToPackage{normalem}{ulem}
\usepackage{ulem}
\usepackage{todonotes}
\usepackage{changes}
\usepackage{lipsum}
\usepackage{bm}
\usepackage{color}
\usepackage{array}
\usepackage{float}
\usepackage{multirow}
\usepackage{calligra}
\usepackage{hyperref}
\usepackage[mathscr]{eucal}

\newcommand{\ii}{\mathrm{i}}

\newcommand{\eref}[1]{Eq.~(\ref{#1})}
\newcommand{\erefs}[1]{Eqs.~(\ref{#1})}

\DeclareMathAlphabet{\mathcalligra}{T1}{calligra}{m}{n}
\DeclareFontShape{T1}{calligra}{m}{n}{<->s*[2.2]callig15}{}

\def\bea{\begin{eqnarray}}
\def\eea{\end{eqnarray}}
\def\ba{\begin{array}}
\def\ea{\end{array}}

\def\la{\langle}
\def\ra{\rangle}

\begin{document}
\title{Propulsion dispersion mediated ordering transition in active particles}
\author{Debraj Dutta}
\email{debraj.dutta@bose.res.in}

\affiliation{S. N. Bose National Centre for Basic Sciences, Kolkata 700106, India}

\author{Urna Basu} 
\affiliation{S. N. Bose National Centre for Basic Sciences, Kolkata 700106, India}

\begin{abstract}
We show that dispersion in propulsion strength qualitatively alters collective behavior of active multi-particle systems interacting via short-range attractive potential, giving rise to novel ordered phases that combine spatial and orientational ordering.  Considering a binary mixture of active Brownian particles with two distinct self-propulsion strengths, we find that, the interplay between interaction range, self-propulsion strengths and the relative numbers of the particles with different propulsion strengths can lead to three different phases, namely, a disordered one, and two ordered ones with partial and complete spatial and orientational ordering.
The partially ordered phase is characterized by formation of a ring-like assembly of the slower particles while the faster particles diffuse randomly. Two concentric rings, comprising faster and slower particles, form in the fully ordered phase. Using the example of a truncated harmonic potential, we analytically characterize the phase boundaries and identify the associated order parameters. Our results demonstrate that propulsion dispersion provides a robust and novel route to collective ordering in attractive active matter. 
\end{abstract}

\maketitle

The inherently nonequilibrium self-propelling motion of active particles leads to a plethora of unusual behaviors both at individual and collective levels~\cite{romanczuk2012,RevModPhys.88.045006,o2022time,fodor2018statistical}. One of the most striking features of collective behaviour of active systems is the emergence of spontaneous ordering such as flocking, swarming, motility-induced phase separation (MIPS) and living crystal formation~\cite{cavagna2014bird, bauerle2020formation, cates2015motility, palacci2013living, kumar2014flocking, cavagna2017dynamic,alert2020physical, malinverno2017endocytic,grossmann2015pattern,szabo2006phase}. While most studies emphasize alignment or long-ranged couplings as the key ingredients for order~\cite{toner1995long,yang2015hydro,zhang2021active,chate2008collective,martin2018collective,linden2019motility,pu2017reentrant,sese2022impact,knevzevic2022collective,vicsek2012collective,chate2008modeling,cavagna2015flocking, liebchen2017collective,cavagna2018physics}, short-range isotropic interactions have recently been recognized as capable of generating complex self-organized states even in the absence of explicit alignment rules~\cite{caprini2023flocking}. Understanding and characterising the mechanism behind such orientational and spatial order remains a central challenge in the physics of active matter. 

A key feature of real active systems--from bacterial colonies to synthetic Janus colloids--is that self-propulsion speeds are rarely uniform~\cite{gough2017biologically, vos2008natural, azhar2001cell,rossine2020eco}. The effect of dispersion in propulsion in active system has been an increasing subject of interest in recent years~\cite{forget2022heterogeneous, lauersdorf2025binary,de2021diversity, de2021active, cates2010arrested,  debnath2020enhanced, rojas2023wetting}. More specifically, it has recently been shown that the presence of dispersion in self-propulsion speed can lead to a range of unusual effects in the collective behaviour of active particles including non-monotonicity in pressure and density~\cite{lauersdorf2025binary} to structural modification to MIPS phase~\cite{saw2024active} and wetting layer formation near the surfaces~\cite{rojas2023wetting}. Such dispersion has also been shown to hinder cluster formation in active systems~\cite{de2021diversity}. From a more microscopic point of view, it has also been observed that dispersion in attractively coupled systems can lead to effective repulsion~\cite{sarkar2025emergent}.  Yet, the role of such propulsion dispersion in shaping collective order has remained largely unexplored. 

In this Letter, we show that diversity in self-propulsion can lead to a set of novel emergent phases with both spatial and orientational orderings. Focusing on a binary mixture of Active Brownian particles interacting via a short-range attractive potential, we show that the interplay between the attractive interaction and diversity in self-propulsion can lead to two different ordered phases---a partially ordered one where slower particles form a ring-like assembly and a fully ordered one where both species of particles form two concentric rings. We analytically characterize the phase boundaries for a simple case of truncated harmonic interaction. The emergence of these phases remains robust for different active dynamics as well as large class of attractive potentials, which we illustrate using numerical simulations with a truncated Lennard-Jones potential.

Our setup consists of a collection of $N$ inertial active Brownian particles, each of mass $m$, confined in a two dimensional torus of size $L \times L$. In the absence of any interaction or external force, each particle self-propels with a constant acceleration along its internal orientation, which itself evolves stochastically. The set-up comprises particles interacting pairwise via some short-range attractive potential $V(r)$, which depends only on the radial distance $r$ between the two particles. The position ${\bm r}_i$ of the $i$-th particle evolves  according to the underdamped Langevin equation,
\begin{align}
    \dot{{\bm r}}_i &= {\bm v}_i, \label{eq:rd_Le}\\
    m\dot{{\bm v}}_i &= -\sum_{j \ne i} \nabla_i V(|{\bm r}_{i}-{\bm r}_{j}|) - \gamma {\bm v}_i + a_i \hat{{\bm n}}_i, \label{eq:vd_Le}
\end{align}
where $\gamma$ is the viscous damping coefficient assumed to be same for all the particles. Moreover, $a_i$ and $\hat{{\bm n}}_i = (\cos\theta_i, \sin\theta_i)$ denote the self-propulsion strength and the internal orientation of the $i$-th particle, respectively. The orientation angles $\{ \theta_i\}$ undergo independent rotational diffusion,
\begin{align}
    \dot{\theta}_i = \sqrt{2D_{R}}\,\eta_i(t), \label{eq:tht_t}
\end{align}
where $\{ \eta_i(t)\}$ are independent Gaussian white noise with time autocorrelation,
\begin{align}
    \langle \eta_i(t)\eta_j(t') \rangle = \delta_{ij}\delta(t - t').
\end{align}
The rotational diffusivity $D_R$ is assumed to be same for all the particles. For simplicity, we consider the case where there are only two distinct values of self-propulsion strength---$N_1$ ($N_2 = N - N_1$) particles move with propulsion strength $a_{1}$ ($a_2$).


\begin{figure}[t]
    \centering
    \includegraphics[width=8.7cm]{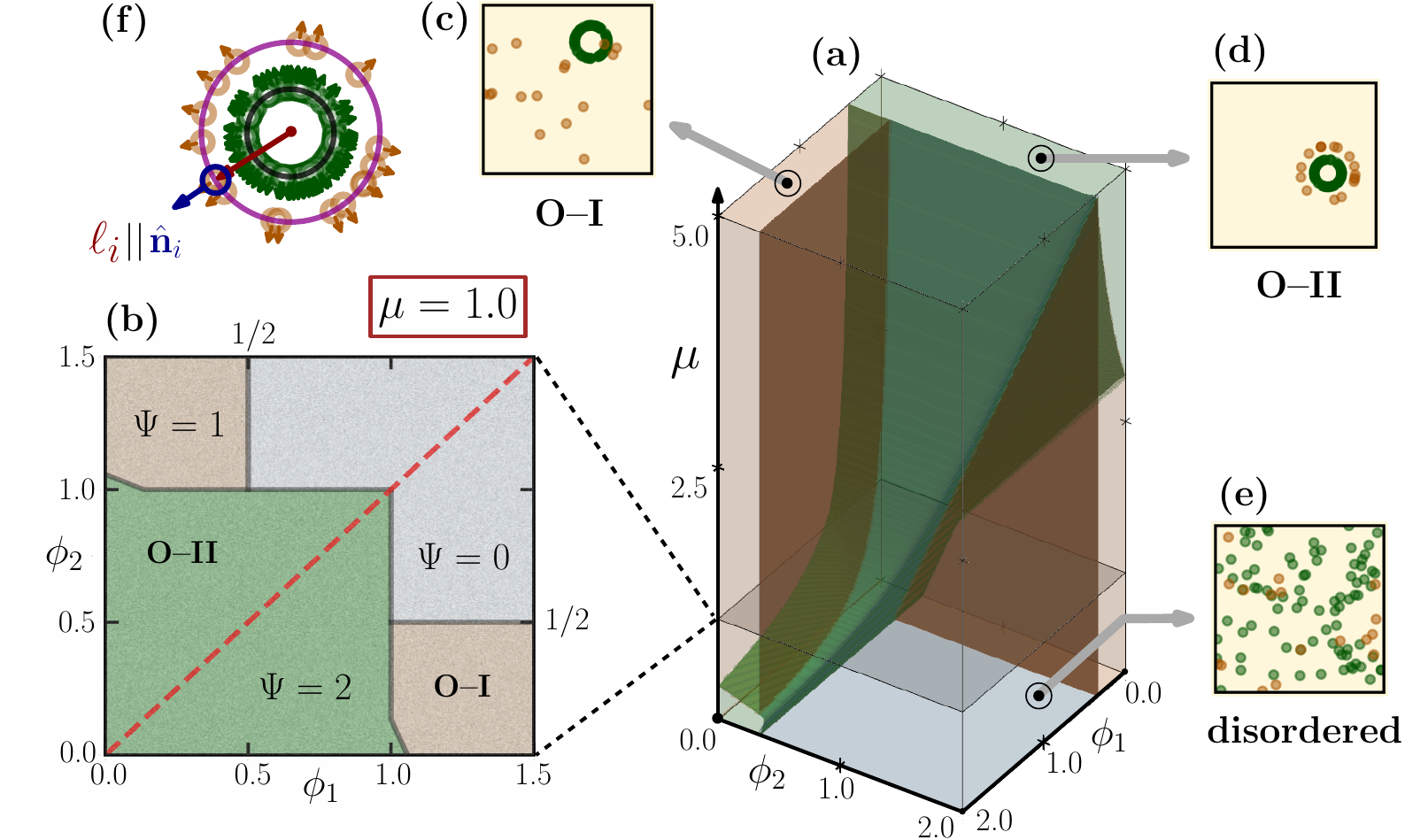}
    \caption{Ordering transitions: 
    (a) Phase diagram in the $(\phi_{1},\phi_{2},\mu)$ space. (b) Phase diagram on the $(\phi_1, \phi_2)$ plane for a fixed $\mu=1$. (c)-(e) show typical configurations in the O-I, O-II and disordered phases, respectively. (f) Schematic diagram showing the orientational ordering.}
    \label{fig:full_phdiag}
\end{figure}


To understand the behaviour of this interacting system we first consider a simple analytically tractable scenario of a truncated harmonic potential,
\begin{align}
    V(r) = \frac{k}{2}\Big[r^2\Theta(r_{0} - r)+r_{0}^2\Theta(r - r_{0})\Big]. \label{eq:V_harm}
\end{align}
We find that the resulting nonequilibrium steady state (NESS), in the strongly active regime $D\ll (\gamma/m, Nk/\gamma)$, exhibits novel emergent phases of spatial and orientational self-organization. In particular,  we show that, depending on the various parameters, the system can be in either of three different phases, namely, disordered, partially ordered or a fully ordered phase. The disordered phase is characterised by independent random motion of all the particles. In the partially ordered phase the slower particles, i.e., the particles with smaller self-propulsion strength self-assemble into a typically circular structure, whereas the faster particles continue to diffuse randomly. Finally, in the fully ordered phase, the faster particles also assemble to form their own ring-like structure outside the slower particle ring, resulting in two concentric circles. Moreover, the spatial ordering is also accompanied by an orientational ordering---the internal orientation vector $\hat{\bm n}_{i}$  of the $i-$th particle aligns with its radial vector from the center of the circles [see Fig.~\ref{fig:full_phdiag}(f)].

To characterize the ring-like structure formation of the $\alpha-$th species, we define the order parameters,
\begin{align}
\kappa_{\alpha}=N_{\alpha}\Big/\sum_{j=1}^{N_{\alpha}}\la |{\bm \ell}_{j}|\ra,~~\text{for}~~\alpha=1,2,\label{eq:OP_K}
\end{align}
where, ${\bm \ell}_{j}={\bm r}_{j}-{\bm R}$, and ${\bm R}=N^{-1}\sum_{j=1}^{N}{\bm r}_{j}$ denotes the position of the center of mass of the particles. In the thermodynamic limit, $\kappa_{\alpha}$, which measures the curvature of the $\alpha-$th ring, vanishes in the disordered phase, and attains a non-zero value when the $\alpha-$th particle species forms a ring. On the other hand, the orientational ordering can be characterized by a different set of order parameters,
\begin{align}
    \Psi_{\alpha}=\frac{1}{N_\alpha}\sum_{j=1}^{N_{\alpha}}\hat{\bm \ell}_{j}\cdot\hat{\bm n}_{j},\label{eq:OP_Psi}
\end{align}
where, $\hat{\bm \ell}_{j}={\bm \ell}_{j}/|{\bm \ell}_{j}|$ is the unit vector along the radial direction. Clearly, $\Psi_{\alpha}$ measures the degree of alignment between the particle orientations and their corresponding radial directions. It attains a value of unity when all particles of species $\alpha$ are exactly aligned with their radial vectors, whereas it vanishes in the disordered phase.

We find that the emergence of the different phases can be most conveniently described by three dimensionless quantities, 
\begin{align}
    \mu=N_{1}/N_{2},\quad\phi_{1} = \frac{a_{1}}{kr_{0}N_{1}},\quad\phi_{2} = \frac{a_{2}}{kr_{0}N_{2}}.\label{eq:dimless}
\end{align}
Physically the system remains invariant under the exchange $a_{1}\leftrightarrow a_{2}$ and $N_{1}\leftrightarrow N_{2}$. Hence, the phase diagram is expected to remain invariant under the transformation $\phi_1\leftrightarrow\phi_2$ and $\mu\leftrightarrow1/ \mu$. Consequently, it suffices to examine the phase diagram in the region $\phi_{2}\ge\mu\phi_{1}$.

In regimes where the propulsion strength of the particle species dominates over their underlying interactions, the collective behavior of the system remains disordered. As either $\phi_{1}$ or $\phi_{2}$ decreases, signatures of ordering begin to emerge. A transition from the disordered phase to a partially ordered phase occurs when the slower particles assemble to form a single ring which we refer to as the \emph{Ordered-I} (O-I) phase. A second transition from the O-I phase to a fully ordered one, referred to as \emph{Ordered-II} (O-II) phase, occurs when the faster particles also organize into a ring, resulting in two concentric circular assemblies. For the truncated harmonic potential, the boundaries separating the three phases can be obtained explicitly. The details of the computations are presented later. Here we summarize the main results.

The disordered and the O-I phases are separated by the surface,
\begin{align}
    \phi_{1}=\frac{1}{2},\quad\Phi(\phi_{1},\mu)\le\phi_{2}<\infty,\label{eq:ph_bndry_1}
\end{align}
where,
\begin{align}
    &\Phi(\phi_{1}, \mu)=\max\Bigg[\frac{1}{2}(1+\mu), \frac{1+\mu(1-\phi_{1})}{1+\mu}\times\cr
    &\Big(\mu+\frac{1}{\pi}\Big[2 \csc^{-1}\frac{\mu\phi_{1}}{1+\mu} - \sin\Big(2 \csc^{-1}\frac{\mu\phi_{1}}{1+\mu}\Big)\Big]\Big)\Bigg].\label{eq:Phi_def}
\end{align}
For fixed $\mu$ and $\phi_{2}$, decreasing $\phi_{1}$ across this surface leads to a transition from the disordered to the O-I phase. Furthermore the O-II is separated from O-I by the surface,
\begin{align}
    \phi_{2} = \Phi(\phi_{1}, \mu),\quad \phi_{1}\le \frac{1}{2},\label{eq:ph_bndry_2}
\end{align}
where, $\Phi(\phi_1,\mu)$ is defined in \eref{eq:Phi_def}. Crossing this surface by decreasing $\phi_{2}$ for fixed $\mu$ and $\phi_{1}$ leads to a transition from the O-I to the O-II phase. Finally, for $\phi_1>1/2$, a direct transition from the disordered phase to the Ordered-II phase can occur along,
\begin{align}
    \phi_{2} = \Phi(\phi_{1}, \mu),\quad \phi_{1}\ge \frac{1}{2}.\label{eq:ph_bndry_3}
\end{align}
The resulting phase diagram, illustrating these regions and transition surfaces, is shown in Fig.~\ref{fig:full_phdiag}(a). 

In the partially ordered phase, the orientational order parameter $\Psi_{1}$ [see \eref{eq:OP_Psi}], associated with the slower particle species, attains a value of unity, while $\Psi_{2}$ for the faster species remains zero. In contrast, in the fully ordered (O-II) phase, both $\Psi_{1}$ and $\Psi_{2}$ become unity, indicating complete orientational alignment within each species. This behavior is illustrated in Fig.~\ref{fig:full_phdiag}(b), where the composite quantity $\Psi = \Psi_{1} + \Psi_{2}$ is plotted in the $(\phi_{1}, \phi_{2})$ plane for a fixed value of $\mu$. Evidently, $\Psi$ takes the values $0$, $1$, and $2$ in the disordered, O-I, and O-II phases, respectively, showing a clear distinction between the different regimes of collective orientational ordering.


\begin{figure}
    \centering
    \includegraphics[width=8.8cm]{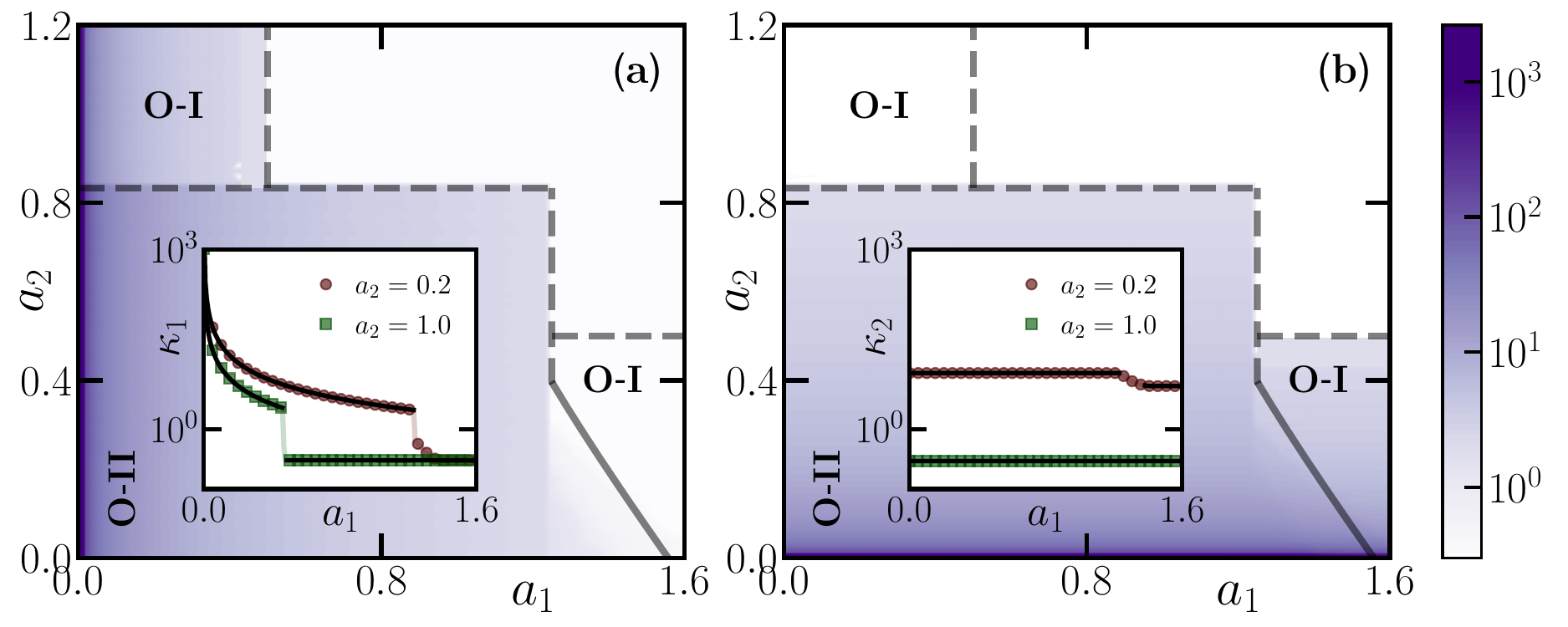}
    \caption{Phase diagram showing the O-I and O-II phases for the truncated harmonic potential: Colour-gradient maps showing the variations of the order parameters $\kappa_{1}$ (a) and $\kappa_{2}$ (b)  on the $a_{1}-a_{2}$ plane, obtained from numerical simulations. The dashed and solid lines indicate the phase boundaries [see \erefs{eq:ph_bndry_1}-\eqref{eq:ph_bndry_3}] for discontinuous and continuous transitions, respectively. The insets show plots of $\kappa_{1,2}$ as functions of $a_1$ for different values of $a_2$. Symbols indicate data obtained from numerical simulations whereas the solid lines correspond to analytical predictions Eqs.~\eqref{eq:OP_KI}–\eqref{eq:OP_KII_2} The other parameters used are $N_1=25,N_{2}=37$, i.e., $\mu=0.67$ and $r_0=1.0, k = 0.04, \gamma=1.0, m=0.01, D=0.01$.}\label{fig:kappa_harmonic}
\end{figure}


The order parameters $\kappa_\alpha$ [see \eref{eq:OP_K}] characterising the spatial ordering, on the other hand, exhibit a rather nontrivial behaviour. In the O-I phase, these order parameters are given by,
\begin{align}
    \kappa_{1} = \frac{1}{r_0 \phi_{1}}=\frac{k N_{1}}{a_{1}}, \quad\kappa_{2} = 0.\label{eq:OP_KI}
\end{align}
In the O-II phase, both order parameters attain finite values. For the slower particles, the order parameter is given explicitly by
\begin{align}
\kappa_{1} = \frac{1}{r_{0} \phi_{1}}\Big(1 + \frac{1}{\mu}\Big)= \frac{k N}{a_1},
\label{eq:OP_KII_1}
\end{align}
while for the faster particles, $\kappa_{2}$ is determined implicitly from the equation [see Appendix~\ref{ap:phases}],
\begin{align}
\phi_{2} = G_{2}\Big(r_{0}\kappa_{2}/2, \mu\Big),~\text{with}~
\left.\frac{d}{dz} G_{2}(z, \mu)\right|_{z = r_{0}\kappa_{2}/2} < 0,
\label{eq:OP_KII_2}
\end{align}
where,
\begin{align}
    G_2(z;\mu) = \begin{cases}
        \displaystyle\frac{\mu}{2z} +\frac{1}{\pi}\Big[\frac{\sin^{-1}z}{z}-\sqrt{1-z^2}\Big],& z\le1,\\[1em]
        \displaystyle\frac{1}{2 z}(\mu + 1),&z\ge1.
    \end{cases}\label{eq:G2_def}
\end{align}
These analytical findings are also supported by extensive numerical simulations [see Appendix~\ref{ap:simulation}]. Figure~\ref{fig:kappa_harmonic} shows behaviour of the order parameters $\kappa_{1}$ and $\kappa_{2}$ in the $(a_{1}, a_{2})$ plane for a fixed value of $\mu$. Note that the two O-I phases shown in the plots correspond to different scenarios. 
In the O-I phase located in the bottom-right corner, $a_{1} > a_{2}$, so that the slower $N_{2}$ particles form the single ring, resulting in $\kappa_{1} = 0$ and $\kappa_{2} > 0$. 
In contrast, the opposite scenario occurs in the O-I phase shown in the top-left corner. The corresponding insets compare the analytical predictions Eqs.~\eqref{eq:OP_KI}–\eqref{eq:OP_KII_2} with results obtained from numerical simulations which show an excellent agreement. 

In the following, we discuss the emergence of the different phases in the context of truncated harmonic potential with some details. We start by recasting the Langevin equations \eqref{eq:rd_Le}-\eqref{eq:vd_Le} in the center of mass frame,
\begin{align}
    m\ddot{\bm{\ell}}_{i} +\gamma \dot{\bm \ell}_{i}&= a_{i}\hat{\bm{n}}_{i}-\sum_{j=1}^{N}\Big[\frac{a_{j}\hat{\bm{n}}_{j}}{N}+{\bm \nabla}_{i} V(|{\bm \ell}_i - {\bm \ell}_j|)\Big]. \label{eq:eom_l}
\end{align}
where ${\bm \ell}_{i}={\bm r}_{i}-{\bm R}$ and $V(r)$ is the truncated harmonic potential defined in \eref{eq:V_harm}. 

In the strongly active regime  the  active time-scale  $D_{R}^{-1}$ is much larger than the other time-scales, namely, the viscous time-scale $m/\gamma$ and the relaxation time-scale in the trap $\gamma/(Nk)$. Thus, in this regime, the 
position vectors $\{{\bm \ell}_{i}\}$ relax long before the particle orientations change appreciably. Eventually the orientation vectors $\hat{\bm{n}}_{i}=(\cos\theta_{i}, \sin\theta_{i})$, reach a uniform steady state in $\theta_{i}\in[0,2\pi]$ for $t\gg D_{R}^{-1}$  [see \eref{eq:tht_t}]. Thus, in this regime, the stationary position distributions can be expressed as,
\begin{align}
    P(\{ {\bm \ell}_i\} ) = \frac{1}{(2\pi)^N}\intop_0^{2\pi} \prod_{i=1}^N d\theta_{i}\,\mathscr{P}(\{ {\bm \ell}_i\}| \{ \theta_i \}),\label{eq:st_dist}
\end{align}
where $\mathscr{P}(\{ {\bm \ell}_i\}| \{ \theta_i \})$ denotes the position distribution of the particles for fixed orientations  $\{ \theta_i \}$. For fixed orientations, the particles evolve deterministically, eventually reaching a configuration which must satisfy the balance equation [see Appendix~\ref{ap:phases}],
\begin{align}
    \sum_{j=1}^{N}\nabla_{i} V(|\bm \ell_{i}-\bm \ell_{j}|)=a_{i}\hat{\bm{n}}_{i}, \label{eq:fbal}
\end{align}
where, for simplicity, we have assumed that the initial orientations of the particles are distributed uniformly in $[0,2\pi]$. 

To obtain the position configuration of the particles following this force balance equation we look for solutions which lead to the different ordered phases of the system. To this end, we consider configurations in which the particles arrange themselves into concentric rings---$N_1$ particles with self-propulsion $a_{1}$ on a circle of radius $\mathscr{R}_{1}$ and $N_2$ particles with self-propulsion $a_{2}$ on circle of radius $\mathscr{R}_{2}$, with $a_{2}>a_{1}$, so that $\mathscr{R}_{2}>\mathscr{R}_1$. For a given interaction cutoff $r_{0}$, we can then write the force balance equations for the slower and faster particle species separately---by summing the attractive interactions acting on a tagged particle from all neighbors within the interaction range and equating the result to the active force. The details are provided in Appendix~\ref{ap:phases}. Here we briefly discuss the main results. We find that there are two distinct cases when these balance equations admit stable solutions. In the first case only the slower particles form a single stable ring while the faster ones remain disordered. This corresponds to the scenario when there is no interaction between the two particle species. The second case represents the fully ordered phase, where both species organize into two concentric rings. This is possible when $r_{0}>\mathscr{R}_{2}+\mathscr{R}_{1}$, i.e., when every particle of one species interacts with all particles of the other. In the following, we analyze these one and two ring configurations separately.

For a single stable ring, the steady-state force balance \eref{eq:fbal}, takes the form [see Appendix~\ref{ap:phases}],
\begin{align}
    \frac{a_{1}}{r_{0}kN_{1}} = G_1\Big(\frac{r_{0}}{ 2\mathscr{R}_{1}}\Big),\label{eq:phi1_G1}
\end{align}
where, $G_1(z) = G_2(z,0)$. The radius $\mathscr{R}_1$ of this ring can be obtained by solving the transcendental \eref{eq:phi1_G1} for fixed $r_0, a_1$ and $N_1$ in a self-consistent manner. Note that, for stable solutions $\mathscr{R}_{1}$ must additionally satisfy $G_1'(r_0 / (2\mathscr{R}_1))<0$. Using this stability condition along with \erefs{eq:phi1_G1}, it is easy to show that the allowed region of phase space supporting a single stable ring is given by,
\begin{align}
    0\le\phi_{1}\equiv\frac{a_1}{r_0 k N_{1}}\le1/2.\label{eq:ph_bd_1}
\end{align}
The value of the spatial order parameter $\kappa_{1}=1/\mathscr{R}_{1}$ in this region is quoted in \eref{eq:OP_KI}. On the other hand, the value of the orientational order parameter $\Psi_{1}$ in this region is always one.

For the case of two stable rings, the force balance equation \eqref{eq:fbal} translates to,
\begin{align}
    \frac{a_1}{k N} = \mathscr{R}_{1} ,\quad \frac{a_2}{k N_{2}} = G_2\Big(\frac{r_0}{k \mathscr{R}_2}; \mu\Big),\label{eq:phi12_G2}
\end{align}
where $G_2$ is defined in \eref{eq:G2_def}. Solving the above equations for fixed values of $r_0, a_1, a_2, N_1$ and $N_2$ gives the radii of the concentric rings and, consequently, the corresponding order parameters $\kappa_i=1/\mathscr{R}_{i}$ [see \erefs{eq:OP_KII_1}-\eqref{eq:OP_KII_2}]. For stable two ring solutions, $\mathscr{R}_{2}$ must additionally satisfy $G_{2}'(r_0 / (2\mathscr{R}_2); \mu)<0$ and the geometric constraint $r_0>\mathscr{R}_1+\mathscr{R}_2$. Equation~\eqref{eq:phi12_G2} together with these constraints give the region of phase space supporting two stable concentric rings,
\begin{align}
    0\le~&\phi_{1}\le\frac{1}{2}\Big(1+\frac{1}{\mu}\Big),~~\text{and}\cr
    0\le~&\phi_{2}\le\max\Bigg[G_2\Big(\frac{1/2}{1-\mathscr{R}_1/r_0};\mu\Big),\frac{1}{2}(1+\mu)\Bigg].\label{eq:ph_bd_2}
\end{align}
Note that in this region both the orientational order parameters $\Psi_{1}$ and $\Psi_{2}$ become unity. Equations \eqref{eq:ph_bd_1} and \eqref{eq:ph_bd_2} specify the part of the phase space in which ordered structures---either one ring or two concentric ring---can exist. 


\begin{figure}
    \centering
    \includegraphics[width=8.8cm]{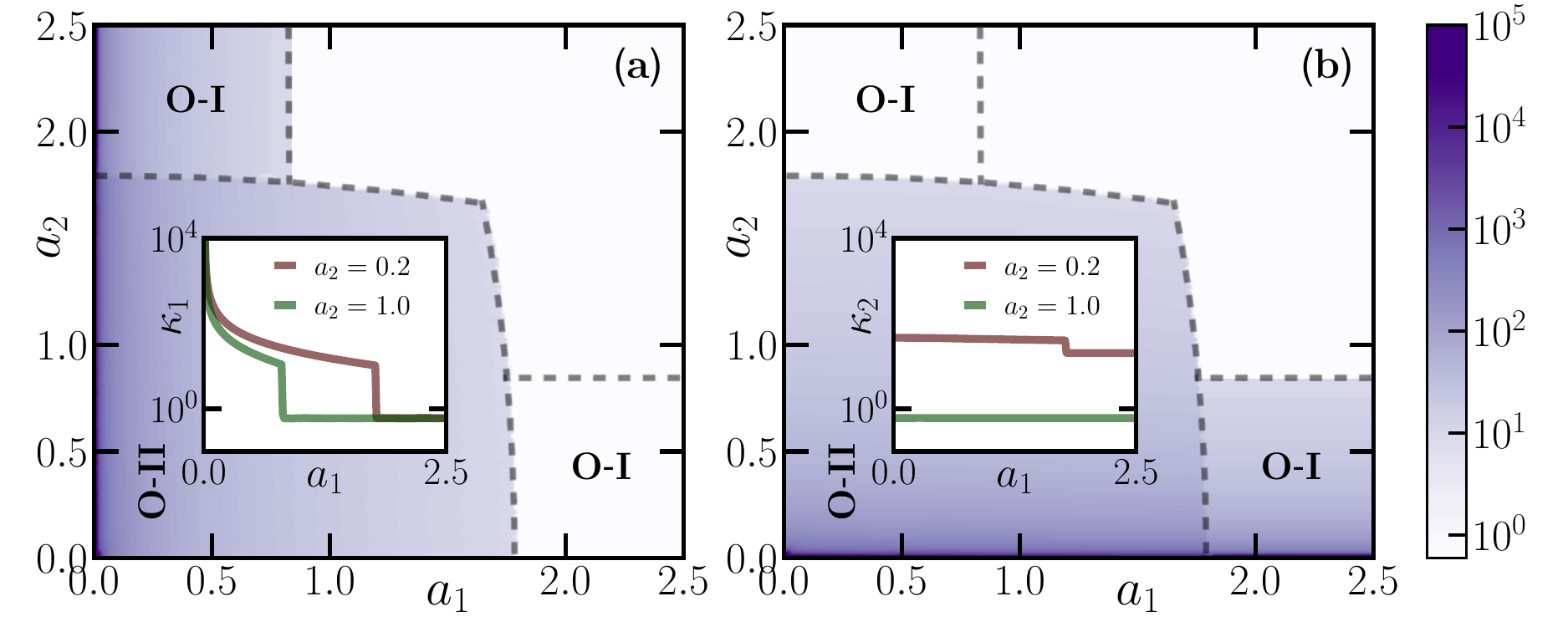}
    \caption{Phase diagram showing the O-I and O-II phases for the truncated Lennard-Jones potential: Colour-gradient maps showing the variations of the order parameters $\kappa_{1}$ (a) and $\kappa_{2}$ (b)  on the $a_{1}-a_{2}$ plane, obtained from numerical simulations. The dashed and solid lines indicate the phase boundaries. The insets show plots of $\kappa_{1,2}$ as functions of $a_1$ for different values of $a_2$ obtained from numerical simulations. The other parameters used are $N_1=25,N_{2}=25$, i.e., $\mu=1.0$ and $r_0=0.2, k = 0.04, \sigma = 5.0, \gamma=1.0, m=0.01, D_R=0.01$.}
    \label{fig:kappa_LJ}
\end{figure}


Interestingly, there exist regions of the phase space where both the single-ring and two-ring solutions satisfy the force-balance and stability criteria. Yet, numerical simulations show no evidence of such coexistence. To determine which configuration is actually realized in these overlapping regions, we compare the corresponding energy costs. The average total energy of the system in each phase can be estimated by considering the contributions from the interaction and the self-propulsion forces. In the overdamped limit it can be shown that [see Appendix~\ref{ap:energy} for details] whenever the single-ring and two-ring phases coexist in the permitted region of phase space, the two-ring configuration is energetically favored.

It is now interesting to examine the behaviour of the order parameters, $\Psi_\alpha$ and $\kappa_\alpha$ across the various phase boundaries. The order parameters change discontinuously when the system transitions from the disordered phase to the O-I phase. In contrast, the transition between the O-I and O-II phases can either be continuous or discontinuous, depending on whether $G_2\Big(1/(2-2\mathscr{R}_1/r_0);\mu\Big)$ is larger or smaller than $(1+\mu)/2$. Physically speaking, the continuous nature of the transition in this case implies that near the transition point the second ring forms with only a fraction of the faster particles whereas the rest of them continue to diffuse freely. In contrast, the single ring always contains all the slower particles in both O-I and O-II phases. This behaviour is illustrated in Fig.~\ref{fig:kappa_harmonic}. 

To investigate the robustness of these ordering transitions for more general short-range interactions, we perform numerical simulations using a truncated Lennard–Jones potential [see Appendix~\ref{ap:simulation} for details]. As shown in Fig.~\ref{fig:kappa_LJ}, the resulting phase diagram in the $a_1$–$a_2$ plane, together with the corresponding behaviors of the order parameters $\kappa_1$ and $\kappa_2$, closely mirrors that obtained for the truncated harmonic interaction. This striking agreement further substantiates our central claim that the emergence of the ordered phases is independent of the specifics of the interaction potential, arising generically due to the interplay between the underlying attraction and propulsion dispersion. It should also be noted that although we used the example of active Brownian particles, our analysis holds true for all other active particle models like run-and-tumble particles and direction reversing active Brownian particles.

In summary, we have shown that presence of self-propulsion dispersion leads to the emergence of novel ordered phases, with both spatial and orientational ordering, in attractive active mixtures. Using the example of truncated harmonic interaction, we analytically characterize the ordered phases and the order parameters. Several interesting questions remain open, including how propulsion dispersion influences MIPS-like phase separation and how broader forms of polydispersity affect collective dynamics. Finally, exploring experimental realizations in synthetic or biological active systems may reveal an even richer spectrum of behaviors driven by propulsion diversity.

\appendix

\section{Computation of the phase boundaries and order parameter $\kappa_{\alpha}$}\label{ap:phases}

In this appendix we provide the details of the computation of the phase boundaries and the order-parameter $\kappa_\alpha$. We start by writing the equations of motion in the centre of mass frame. Using \erefs{eq:rd_Le} and \eqref{eq:vd_Le}, we find the Langevin equation for the centre of mass ${\bm R}=N^{-1}\sum_{i=1}^{N}{\bm r}_{i}$,
\begin{align}
    m \ddot{\bm{R}}=-\frac{\gamma}{N}\sum_{i=1}^{N}\dot{\bm{r}}_{i}+\frac{1}{N}\sum_{i=1}^{N} a_{i} \hat{\bm{n}}_{i}.
\end{align}
Consequently, the centre of mass coordinates of the $i$-th particle $\bm {\ell}_{i}={\bm r}_{i}-{\bm R}$ evolve via \eref{eq:eom_l}. The orientational distribution of the particles is uniform in the steady state and for large $N$, we have, to leading order, $N^{-1}\sum_{j=1}^{N}\big(a_{i}\hat{\bm{n}}_{i}-a_{j}\hat{\bm{n}}_{j})\simeq a_{i}\hat{\bm{n}}_{i}$ in \eref{eq:eom_l}. In the strongly active regime, i.e., when $D_{R}^{-1}\gg \big(m/\gamma,~\gamma/(Nk)\big)$, the timescale separation allows the stationary distribution to be expressed as \eqref{eq:st_dist}. Thus we need to compute $\mathscr{P}(\{ \ell_i\}| \{ \theta_i \})$ which can be obtained by setting $\dot{\bm \ell}$ and $\ddot{\bm \ell}$ to zero in \eref{eq:eom_l}. This leads to the force balance equation \eqref{eq:fbal}. For uniformly distributed initial orientations, it is convenient to recast \eref{eq:fbal} in the matrix form,
\begin{align}
    K \psi = F, \label{eq:mat}
\end{align}
where, $K$ is the $N\times N$ interaction matrix with off-diagonal elements,
\begin{align}
    K_{ij}=\begin{cases}
        -1,&|\bm{\ell}_{i}-\bm{\ell}_{j}|\le r_0, \cr
        0,&|\bm{\ell}_{i}-\bm{\ell}_{j}|> r_0,
    \end{cases}\label{eq:Kdef}
\end{align}
and diagonal elements $K_{ii}=-\sum_{j\ne i}^{N}K_{ij}$. Moreover, we have defined the column vectors of size $N$ ,
\begin{align}
    \psi = \begin{pmatrix}
                \bm{\ell}_1\\ {\bm \ell_2}\\ \vdots\\ {\bm \ell_{N}}
            \end{pmatrix},\, 
    F = \frac{1}{k}\begin{pmatrix}
            a_{1} \mathscr{F}_{1}\\ 
            a_{2} \mathscr{F}_{2}    
        \end{pmatrix},
        \,\text{with}~
    \mathscr{F}_\alpha = \begin{pmatrix}
                        1\\ e^{2\pi\ii/ N_\alpha}\\ \vdots\\e^{2\pi\ii (N_\alpha-1)/ N_\alpha}
                    \end{pmatrix}.
\end{align}
The $j$-th entry of the vector $\mathscr{F}_\alpha$ encodes the self-propulsion direction of the $j$-th particle of the $\alpha$-th species. It should be noted that $\mathscr{F}_{\alpha}$ denotes the eigenvector of the $N_\alpha\times N_\alpha$ cyclic permutation matrix $P_\alpha$ and satisfies the eigenvalue equation,
\begin{align}
    P_\alpha \mathscr{F}_{\alpha}=e^{2\pi\ii/ N_\alpha} \mathscr{F}_\alpha.\label{eq:perMat}
\end{align}
 
\begin{figure}
    \centering
    \includegraphics[width=9cm]{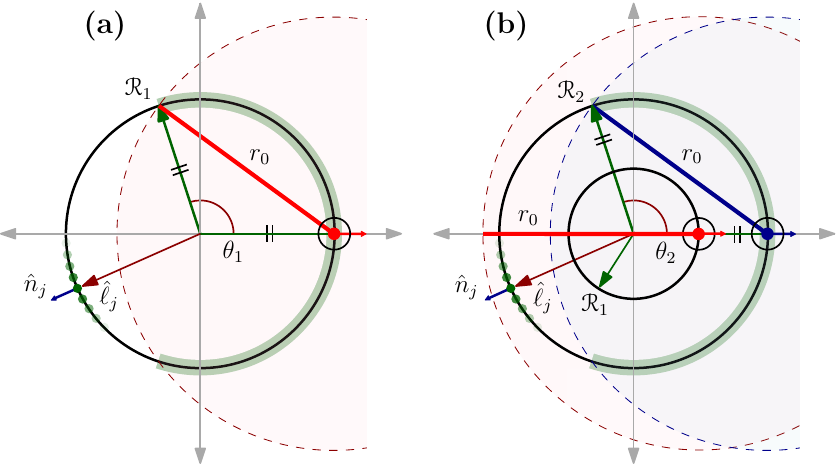}
    \caption{Schematic diagram showing (a) single ring configuration  and (b) two concentric ring configuration.}
    \label{fig:schem}
\end{figure}

The dependence of the matrix $K$ on position vector $\psi$ (through $\{\bm \ell_{j}\}$) makes \eref{eq:mat} inherently non-linear. To understand the behaviour of the ordered phases,  we look for solutions which lead to the formation of concentric rings. To this end, we propose the ansatz,
\begin{align}
    \psi=\left[\mathscr{R}_{1} \mathscr{F}_{1} \atop\mathscr{R}_{2} \mathscr{F}_{2}\right],\quad K= \begin{bmatrix}
         K^{(1)}         &  \tilde {K} \\
         \tilde{K}^T               & K^{(2)}  \\
    \end{bmatrix},\label{eq:ansatz}
\end{align}
where, $\mathscr{R}_\alpha$ denotes the radius of the $\alpha$-th ring. Here we have assumed that $N_\alpha$ particles of $\alpha$ species are distributed uniformly on ring of radius $\mathscr{R}_\alpha$ [see Fig.~\ref{fig:schem}]. The diagonal blocks $K^{(\alpha)}$ are $N_\alpha\times N_\alpha$ square matrices and encode the intra-species interaction between particles of the $\alpha$-th species whereas the off-diagonal $N_1\times N_{2}$ block $\tilde{K}$ denotes inter-species interaction. Substituting \eref{eq:ansatz} in \eref{eq:mat} results in a pair of matrix equations, 
\begin{align}
    \mathscr{R}_{1}K^{(1)}\mathscr{F}_{1}+\mathscr{R}_{2}\tilde{K}\mathscr{F}_{2}&=\frac{a_{1}}{k}\mathscr{F}_{1},\label{eq:mat_eq_1}\\
    \mathscr{R}_{1}\tilde{K}^T\mathscr{F}_{1}+\mathscr{R}_{2}K^{(2)}\mathscr{F}_{2}&=\frac{a_{2}}{k}\mathscr{F}_{2}.\label{eq:mat_eq_2}
\end{align}
For non-zero values of ($\mathscr{R}_{1}$, $\mathscr{R}_{2}$) and arbitrary ($N_{1}, N_2$), the above equations imply,
\begin{align}
    K^{(1)}\mathscr{F}_{1}=\lambda_{1}\mathscr{F}_{1},&~~K^{(2)}\mathscr{F}_{2}=\lambda_{2}\mathscr{F}_{2},\label{eq:K_ij_1}\\
    \tilde K\mathscr{F}_{2}=\tilde\lambda_{1}\mathscr{F}_{1},&~~\tilde K^{T}\mathscr{F}_{1}=\tilde \lambda_{2}\mathscr{F}_{2},\label{eq:K_ij_2}
\end{align}
where $\lambda_{1,2}$ and $\tilde{\lambda}_{1,2}$ are scalar numbers which need to be determined. Eigenvalue \eref{eq:perMat} together with \erefs{eq:K_ij_1} and \eqref{eq:K_ij_2} suggest that the diagonal blocks $K^{(\alpha)}$ commute with $P_\alpha$, whereas the off-diagonal blocks satisfy the relations,
\begin{align}
    P_{1} \tilde{K} = \tilde{K} P_{2}, \quad P_{2} \tilde{K}^T = \tilde{K}^T P_{1}. \label{eq:offd}
\end{align}
Using \erefs{eq:K_ij_1}-\eqref{eq:offd}, it is now easy to show that the eigenvalues $\lambda_\alpha$ are given by,
\begin{align}
    \lambda_{\alpha}=\sum_{j=0}^{N_{\alpha}-1} K^{(\alpha)}_{0j}e^{2\pi\ii j/N_\alpha},\label{eq:eig}
\end{align}
whereas the off-diagonal blocks satisfy the constraints,
\begin{align}
    \tilde K\mathscr{F}_{2}=0,~~\tilde K^{T}\mathscr{F}_{1}=0.\label{eq:offd_constr}
\end{align}
Consequently, \erefs{eq:mat_eq_1} and \eqref{eq:mat_eq_2} now reduce to,
\begin{align}
    \mathscr{R}_{\alpha}\lambda_{\alpha} = \frac{a_{\alpha}}{k},\quad\alpha=1,2.\label{eq:mat_eq_final}
\end{align}
To solve the above equations for $\mathscr{R}_{\alpha}$, we must explicitly compute the eigenvalues $\lambda_{\alpha}$ subject to the constraint \eqref{eq:offd_constr}. Clearly, there are two distinct scenarios satisfying \eref{eq:offd_constr}--(a) when $\tilde{K}_{ij}=0$, with no interaction between the two kinds of particles, and (b) when $\tilde{K}_{ij}=-1$ so that every particle of first kind is visible to every particle of second kind and vice versa. The former case represents the scenario when a single stable ring is formed by the slower particles with the faster particles diffusing randomly in the box. The second scenario corresponds to the case when the remaining faster particles also join to form two concentric rings. We consider the cases of one and two stable rings separately.

We start with the case of a single stable ring formed by the slower particles, i.e., when $\tilde{K}_{ij}=0$. Also with the faster particles diffusing randomly, we additionally have $K^{(2)}_{ij}=0$. It follows immediately from \eref{eq:eig} that $\lambda_{2}=0$, which on substituting in \eref{eq:mat_eq_final} yields the radius $\mathscr{R}_{2}=\infty$. Consequently in this phase, the order parameters $\kappa_{2}$ vanishes. To compute $\lambda_{1}$ we consider the geometric construction shown in Fig.~\ref{fig:schem}(a), consisting of a single ring of radius $\mathscr{R}_{1}$ formed by $N_{1}$ uniformly distributed particles, each with self-propulsion strength $a_{1}$. The effective interaction region of a tagged particle is marked by a circle of radius $r_{0}$. It is now convenient to introduce the dimensionless quantity $z_{1}=r_{0} / (2\mathscr{R}_{1})$ defined on the interval $[0, \infty)$. Depending on whether $z_{1}<1$ or $z_{1}\ge1$ two branches of solutions can exist. For $z_{1}\ge1$, the tagged particle can see every other particle on the ring, and $K^{(1)}$ is simply given by,
\begin{align}
    K^{(1)}_{ij} = f_{K}(N_{1},i,j)\equiv\begin{cases}
        N_1-1,&i=j,\\
        -1,&i\ne j,
    \end{cases}
\end{align}
where we have defined the function $f_{K}$ for future convenience. On the other hand, for $z_{1}<1$, only a fraction of the total particles $N_{1}'$\,$(< N_1)$ on the ring are visible to the tagged particle. Using the angle $\theta_1=2\sin^{-1}z_{1}$, marked in Fig.~\ref{fig:schem}(a), $N'_1$ can be computed using the relation, $N_1'=N_1\theta_1/\pi$. Therefore, in this case, $K^{(1)}_{00}=N'_{1}-1$  and, for $j>0$
\begin{align}
    K^{(1)}_{0j}=\begin{cases}
        -1,&\min\Bigg[\displaystyle\frac{2\pi j}{N_{1}},~2\pi-\displaystyle\frac{2\pi j}{N_{1}}\Bigg]<\theta_{1},\\
        0,&\text{otherwise}.
    \end{cases}
\end{align}
It is now straightforward to compute the eigenvalue $\lambda_{1}$ using \eref{eq:eig}, which in the thermodynamic limit $N_1 \gg 1$, yields,
\begin{align}
    \lambda_{1} &= \begin{cases}
        \displaystyle\frac{N_{1}}{\pi}(\theta_1 - \sin \theta_1),&z_{1}\le1,\\
        N_1,&z_{1}>1. 
    \end{cases}
\end{align}
Substituting $\lambda_{1}$ in \eref{eq:mat_eq_final}, we arrive at the expression quoted in \eref{eq:phi1_G1}.


Next we consider the case of two concentric stable rings, i.e., when $\tilde{K}_{ij}=-1$ and every particle of first kind is visible to every particle of second kind. This is possible when $r_{0} \ge \mathscr{R}_{1} + \mathscr{R}_{2}$ as is illustrated by the geometric construction in Fig.~\ref{fig:schem}(b). The tagged particle on the smaller ring can also see every other particle on the smaller ring, and hence the elements of $K^{(1)}$ are thus given by $K^{(1)}_{ij}=f_{K}(N,i,j)$. As before, we introduce the dimensionless parameters $z_{i}=r_{0} / (2 \mathscr{R}_{i})$, defined on the intervals,
\begin{align}
    1\le z_{1}<\infty,\quad\text{and},\quad\frac{z_{1}}{2z_{1}-1}\le z_{2}<\infty.\label{eq:z_bound}
\end{align}
In deducing the above bounds we have used the inequalities $\mathscr{R}_{2}\ge\mathscr{R}_{1}$ and $r_{0}\ge\mathscr{R}_{1}+\mathscr{R}_{2}$. Again, two branches of solutions can exist depending on whether $z_{2}\le1$ or $z_{2}>1$. For $z_{2}>1$, each particle on the outer ring interacts with all other particles, implying, $K^{(2)}_{ij} = f_{K}(N,i,j)$. For $z_{2}\le1$, a fraction of the total particles $N'_{2}$($<N_{2}$) on the outer ring are visible to the tagged particle on the outer ring and is given by, $N'_{2} = N_{2}\theta_2/\pi$, where $\theta_{2}=2\sin^{-1}z_{2}$. Accordingly, $K^{(2)}_{00}=N'_2-1$ and,
\begin{align}
    K^{(2)}_{0j}=\begin{cases}
        -1& \text{for}\, \, \min\Bigg[\displaystyle\frac{2\pi j}{N_{2}},~2\pi-\displaystyle\frac{2\pi j}{N_{2}}\Bigg]<\theta_{2},\\
        0 &\text{otherwise},
    \end{cases}
\end{align}
for $j>0$. Computing the eigenvalues $\lambda_1$ and $\lambda_2$ using \eref{eq:eig} and substituting in \eref{eq:mat_eq_final} yields \eref{eq:phi12_G2} quoted in the main text.

\section{Computation of energetics of the different phases}\label{ap:energy}

In this section, we estimate the energetic costs associated with the different phases of ordering. To this end, we construct an effective energy functional consisting of distinct contributions from kinetic, interaction, and active forces. In the overdamped limit, particles that diffuse freely in the box move with an almost constant speed $a_{\alpha}/\gamma$. Their contribution to the total energy is therefore purely kinetic and equals $m a_{\alpha}^{2}/(2\gamma^{2})$  per particle of species $\alpha$. The remaining particles form one or more stable ring structures and contribute to the potential part of the energy, which consists of an attractive interaction term and a repulsive active term.

In the disordered phase, all particles diffuse freely and there is no potential energy contribution. The total energy is thus purely kinetic and given by
\begin{align}
    E_{\mathrm{D}} = \sum_{\alpha}\frac{N_{\alpha}}{2}\, m \frac{a_{\alpha}^{2}}{\gamma^{2}}.
\end{align}
In the O-I phase, the faster particles (with propulsion $a_2$) diffuse freely and contribute a kinetic energy $N_{2} m a_{2}^{2}/(2\gamma^{2})$, while the slower particles (with propulsion $a_1$) form a single stable ring. The latter contribute only to the potential energy. Following a procedure analogous to Appendix~\ref{ap:phases}, the attractive interaction energy is obtained by summing the pairwise interaction energies over all particles in the ring. In addition, the active contribution to the potential energy is given by $-a_{1} N_{1} \mathscr{R}_{1}$. Using the expression for $\mathscr{R}_{1}$ from \eref{eq:phi1_G1}, the total energy of the O-I phase becomes
\begin{align}
    E_{\mathrm{I}} = N_{2}\frac{m a_{2}^{2}}{2\gamma^{2}} - \frac{a_{1}^{2}}{2k}.
\end{align}
In the fully ordered phase (O-II), both species form stable rings and the energy is purely potential. Depending on whether $z_{2} \le 1$ or $z_{2} > 1$, corresponding to the two branches of solutions, the interaction energy takes the form
\begin{align}
    U_{K} =
    \begin{cases}
        U_{1} + \tilde{U} + \dfrac{U_{2}}{\pi}\bigl[\theta_{0} - \sin\theta_{0}\bigr], & z_{2} \le 1, \\
        U_{1} + \tilde{U} + U_{2}, & z_{2} > 1,
    \end{cases}
\end{align}
where $U_{\alpha}$ denotes the intra-species interaction energy and $\tilde{U}$ the inter-species contribution, given explicitly by
\begin{align}
    U_{\alpha} = \frac{1}{2}k N_{\alpha}^{2} \mathscr{R}_{\alpha}^{2},
    \quad
    \tilde{U} = \frac{1}{2}k N_{1}N_{2}\bigl(\mathscr{R}_{1}^{2} + \mathscr{R}_{2}^{2}\bigr).
\end{align}
Including the active contributions, $-\sum_{\alpha} a_{\alpha} N_{\alpha} \mathscr{R}_{\alpha}$, the total energy of the O-II phase is
\begin{align}
    E_{\mathrm{II}} = U_{K} - a_{1} N_{1} \mathscr{R}_{1} - a_{2} N_{2} \mathscr{R}_{2}.
\end{align}
Substituting the expressions for $\mathscr{R}_{1}$ and $\mathscr{R}_{2}$ from \eqref{eq:phi12_G2}, it follows that the energies satisfy the ordering,
\begin{align}
    E_{\mathrm{D}} \le E_{\mathrm{I}} \le E_{\mathrm{II}}.
\end{align}
The above expression is used to identify the most stable phase in regions of parameter space where the force-balance criterion allows multiple phases to coexist.

\section{Simulation Details}\label{ap:simulation}
In this appendix, we provide the details of numerical simulations. The Langevin \erefs{eq:rd_Le}- \eqref{eq:tht_t} are integrated using the Euler-Maruyama update scheme~\cite{higham2001algorithmic}. To this end, we write using \erefs{eq:rd_Le}-\eqref{eq:tht_t} the time-discretised update rules for ${\bm r}_{i}(t)$, ${\bm v}_{i}(t)$ and $\theta_{i}(t)$,
\begin{align}
    {\bm r}_{i}(t+\Delta t) &= {\bm r}_{i}(t) + {\bm v}_{i}(t)\Delta t,\cr
    {\bm v}_{i}(t+\Delta t) &= {\bm v}_{i}(t) + \Big[\frac{a_{i}}{m}\hat{\bm n}_{i}(t)\cr &-\frac{1}{m}\sum_{j\neq i}\nabla_{i}V(|{\bm r}_{i} - {\bm r}_{j}|) \Big]\Delta t.
\end{align}
and,
\begin{align}
    \theta_{i}(t+\Delta t) = \theta_{i}(t) + \sqrt{2D\Delta t}~\xi_{i},
\end{align}
where $\xi_{i}$s are independent random numbers drawn from the normal distribution $\mathcal{N}(0,1)$. Simulations are carried out using two types of attractive interactions: (i) a truncated harmonic potential defined in \eref{eq:V_harm}, and (ii) a shifted Lennard–Jones potential,
\begin{align}
    V(r) = 4 k\Big[\Big(\frac{\sigma}{r+r_s}\Big)^{12} - \Big(\frac{\sigma}{r+r_s}\Big)^{6}\Big],
\end{align}
for $r<r_0$ and zero otherwise, with the shift set to $r_s=2\sigma^{1/6}$. To ensure periodic boundary conditions, the computation of the forces $\nabla_{i} V(r_{ij})$ is carried out using the minimal image convention~\cite{frenkel2023understanding}. The system is first allowed to evolve until it reaches a stationary state. Measurements of observables are then performed and averaged within this steady-state regime.

The computation of the ordered parameters $\kappa_\alpha$ [see~\eref{eq:OP_K}] and $\Psi_\alpha$ [see~\eref{eq:OP_Psi}] requires the measurement of the centre of mass, ${\bm R}$, of the system. In a periodic domain of size $L\times L$, the centre of mass is computed by first embedding every particle coordinates onto the unit circle to avoid boundary discontinuities: for $j=1,\dots,N$ define,
\begin{align}
z_{x,j} \;=\;\exp\!\biggl(\frac{\ii\,2\pi\,x_j}{L}\biggr),
\quad
z_{y,j} \;=\;\exp\!\biggl(\frac{\ii\,2\pi\,y_j}{L}\biggr).
\end{align}
Then the mean complex positions are given by,
\begin{align}
\bar z_x \;=\;\frac{1}{N}\sum_{j=1}^N z_{x,j},
\quad
\bar z_y \;=\;\frac{1}{N}\sum_{j=1}^N z_{y,j},
\end{align}
whose arguments are the wrapped‐mean angles on each ring.  Finally, inverting back to Cartesian coordinates yields,
\begin{align}
X =\frac{L}{2\pi}\,\arg(\bar z_x)\bmod L,
Y =\frac{L}{2\pi}\,\arg(\bar z_y)\bmod L,
\end{align}
so that ${\bm R}=(X, Y)\in[0,L)\times[0,L)$ correctly reflects the toroidal topology without artificial jumps at the boundaries.

\bibliographystyle{apsrev4-2}
\bibliography{cite.bib}
\end{document}